\documentclass[traditabstract]{aa} % for the abstract without structuration 
\usepackage{graphicx}
\usepackage{txfonts}
\begin{document}
   \title{The $SOPHIE$ search for northern extrasolar planets\thanks{Based on observations made with 
the SOPHIE spectrograph on the 1.93-m telescope at Observatoire de Haute-Provence (CNRS/OAMP), 
France (program 07A.PNP.CONS). }}

   \subtitle{III. A Jupiter-mass companion around HD\,109\,246}

\author{
 I. Boisse\inst{1}
\and A. Eggenberger\inst{2}  
\and N.C. Santos\inst{3}
\and C. Lovis\inst{4}
\and F. Bouchy\inst{1,5}
\and G. H\'ebrard\inst{1}
\and L. Arnold\inst{5}
\and X. Bonfils\inst{2,4}
\and X. Delfosse\inst{2}
\and M. Desort\inst{2}
\and R.F. D\'iaz\inst{1}
\and D. Ehrenreich\inst{2}
\and T. Forveille\inst{2} 
\and A. Gallenne\inst{6}
\and A.M. Lagrange\inst{2}
\and C. Moutou\inst{7} 
\and S. Udry\inst{4}
\and F. Pepe\inst{4}
\and C. Perrier\inst{2} 
\and S. Perruchot\inst{5}
\and F. Pont\inst{8}
\and D. Queloz\inst{4}
\and A. Santerne\inst{7}
\and D. S\'egransan\inst{4}
\and A. Vidal-Madjar\inst{1}
}

\institute{
Institut d'Astrophysique de Paris, UMR7095 CNRS, Universit\'e Pierre \& Marie Curie, 
98bis Bd Arago, 75014 Paris, France
\and 
Universit\'e Joseph Fourier -- Grenoble 1 / CNRS,  Laboratoire d'Astrophysique de 
Grenoble (UMR 5571), BP 53, 38041 Grenoble Cedex 9, France
\and
Centro de Astrof\'isica, Universidade do Porto, Rua das Estrelas, 4150-762 Porto, Portugal
\and
Observatoire de Gen\`eve, Universit\'e de Gen\`eve, 51 Ch. des Maillettes, 1290 Sauverny, 
Switzerland
\and
Observatoire de Haute Provence, CNRS/OAMP, 04870 St Michel l'Observatoire, France
\and
European Southern Observatory, Casilla 19001, Santiago 19, Chile
\and 
Laboratoire d'Astrophysique de Marseille, Universit\'e de Provence \& CNRS, 38 rue Fr\'ed\'eric Joliot-Curie, 
13388 Marseille cedex 13, France
\and
School of Physics, University of Exeter, Exeter, EX4 4QL, UK 
}
   \date{Received ; accepted }

% \abstract{}{}{}{}{} 
% 5 {} token are mandatory
 
  \abstract
   {We report the detection of a Jupiter-mass planet discovered 
   with the SOPHIE spectrograph mounted on the 1.93-m telescope at the 
   Haute-Provence Observatory. The new planet orbits HD\,109\,246, a G0V star
   slightly more metallic than the Sun. HD\,109\,246b has a minimum mass of 
   0.77 M$_{\rm Jup}$, an orbital period of 68 days, and an eccentricity of 
   0.12. It is placed in a sparsely populated region of 
   the period distribution of extrasolar planets. We also present a correction method for the so-called seeing effect that affects the SOPHIE radial velocities. We complement this discovery 
   announcement with a description of some calibrations that are implemented 
   in the SOPHIE automatic reduction pipeline. These calibrations allow the derivation of 
   the photon-noise radial velocity uncertainty and some useful stellar properties 
   ($v \sin i$, [Fe/H], $\log$\,$R'_\mathrm{HK}$) directly from the SOPHIE data.
    }


\titlerunning{Jupiter-mass companion around HD\,109\,246}
\authorrunning{I. Boisse et al.}

\maketitle
%
%________________________________________________________________

\section{Introduction}

Fifteen years after the first discovery of a planet orbiting another 
solar-type star (Mayor \& Queloz 1995), about 450 extrasolar planets 
have been announced. Most of them were detected with the same historical 
technique: high-precision radial velocimetry. Although several methods 
now allow the routine detection of extrasolar planets, high-precision 
radial velocimetry will remain at the forefront of exoplanet science in the 
coming years. Indeed, due to continuous efforts to improve the instrumental 
sensitivity (e.g. Mayor et al. 2009) and to overcome the limitation of stellar activity 
(e.g. Udry et al. 2006), radial velocimetry is a particularly efficient technique to 
detect low-mass planets (e.g. Mayor et al. 2009). Furthermore, radial-velocity (RV) 
measurements are needed to establish the true nature of transiting planetary 
candidates, to derive their mass, and to measure spin-orbit alignment angles 
via the Rossiter-MacLaughlin effect (e.g. Queloz et al. 2000; Ohta et al. 2005; Winn 2010; Triaud et al. 2010). The numerous projects of  
future high-resolution spectrographs operating either at infrared 
(SPIRou, UPF, SIMPLE) or at visible wavelengths (ESPRESSO, CODEX) denote the 
field development and illustrate %confirm 
the importance of RV measurements in 
exoplanetary science.

The high-precision SOPHIE spectrograph (Perruchot et al. 2008; Bouchy et al. 2009b) has replaced ELODIE (Baranne et al. 1996; Queloz et al. 1998b) at the Cassegrain focus of the 1.93-m telescope at the Haute-Provence 
Observatory (OHP, France). Opened to the community since October 2006, SOPHIE has led to many discoveries in various research fields, including the 
follow-up of transiting planet candidates from the SuperWASP 
(e.g. Cameron et al. 2007), HAT (Bakos et al. 2008) and CoRoT 
(e.g. Bouchy et al. 2008) surveys; asteroseismology 
(Mosser et al. 2008); and the dynamics of binary stars 
(Albrecht et al. 2009). In late-2006, the SOPHIE consortium initiated a large program to search for and characterize extrasolar planets (Bouchy et al. 2009b). This comprehensive survey is both a continuation and an extension 
of the planet-search programs carried out with the ELODIE spectrograph (e.g. Mayor \& Queloz 1995; Delfosse et al. 1998;  Perrier et al. 2003; Galland et al. 2005; Naef et al. 2005; Da Silva et al. 2006).

HD\,109\,246 was observed as part of the second subprogram from the
SOPHIE consortium. This subprogram comprises two parts: a survey for giant 
planets in a volume-limited sample of $\sim$2000 FGK dwarfs, and a follow-up of known
transiting giant planets (Bouchy et al. 2009b). Previous discoveries 
from the main survey include the detection of a massive planet (or light brown 
dwarf) around HD\,16760 (Bouchy et al. 2009b) and the detection of a two-planet 
system around HD\,9446 (H\'ebrard et al. 2010a). The follow-up part allowed the 
observation of the spectroscopic transits (Rossiter-McLaughlin effect) of three 
planets and revealed the two first cases of spin-orbit misalignement 
(Loeillet et al. 2008; H\'ebrard et al. 2008; Moutou et al. 2009; Pont et al. 2009; H\'ebrard et al. 2010b).

Here, we report the discovery of a Jupiter-mass planet orbiting HD\,109\,246. The observations are presented in Sect.~2 and the characteristics of the host 
star are described in Sect.~3. In Sect.~4 we derive the best-fit Keplerian 
model to the velocities of HD\,109\,246 and correct for an instrumental effect on radial velocities due to seeing variation. We summarize and discuss our results in Sect.~5. 
In the Appendices, we describe the calibrations implemented in the SOPHIE 
automatic reduction pipeline to derive the photon-noise RV uncertainty (\ref{uncertainty}), 
the projected rotational velocity $v \sin i$ (\ref{vsini}), the metallicity index [Fe/H] (\ref{metallicity}), 
and the activity index $\log$\,$R'_\mathrm{HK}$ (\ref{activityIndex}). 
%We end the paper by presenting a method %for correcting an instrumental effect known as the seeing effect (D).
%to  (\ref{spectrographIllumination}).

%_________________________________________________________________

\section{Radial-velocity measurements}

We observed HD\,109\,246 with the SOPHIE spectrograph from January 2007 to February 
2010. The measurements were performed in high-resolution mode (resolution power 
of $\Delta\lambda/\lambda\,\approx\,75\,000$) and at constant signal-to-noise ratio 
(SNR variations $\sigma$$_{SNR}$$\simeq$8\%). Exposure times ranged from 270 to 
1200\,s, following the variations of seeing and sky transparency at OHP. We used the 
objAB observing mode with one of SOPHIE's fiber recording the starlight and the other fiber observing the sky to estimate the 
background contamination (e.g. moonlight). % while 
%the other was put on the sky to estimate the background contamination (e.g. moonlight). 
Both fibers were fed with a Thorium-Argon lamp every $\sim$\,2\,hour for %simultaneous 
wavelength calibration. SOPHIE is environmentally stabilized and the intrinsic 
drift of the spectrograph (mainly due to small variations of pressure and
temperature) is less than 3 ms$^{-1}$ per hour. Each exposure was corrected for the 
instrumental drift by interpolating the measured intrinsic drift over the 
$\sim$\,2\,hour period. This method allows a precision of $\sim$\,2\,ms$^{-1}$ 
on the correction of the instrumental drift as determined empirically comparing the interpolated drift with the measured one from observations made with simultaneous wavelength calibration.

 We used the SOPHIE automatic pipeline (Bouchy et al. 2009b) to extract the spectra from the detector images and to cross-correlate them with a G2-type mask derived from the Sun spectra.  %In the data reduction process, we 
%identified three spectra as unusable for the subsequent analysis (Sects. 2 and 3). 
We obtained the radial velocities by fitting each resulting cross-correlation function (CCF) with a 
Gaussian (Baranne et al. 1996; Pepe et al. 2002). 
At this stage we rejected three spectra. 
Two of these spectra had a low signal-to-noise ratio (SNR(550nm)$\leqslant$30), which 
rendered the correction for the CCD charge transfer inefficiency (CTI) less effective 
(Bouchy et al. 2009a). The third spectrum was contaminated by moonlight.

The final RV data set for HD\,109\,246 comprises 58 measurements with a typical SNR of 
47 (per pixel at 550 nm). This leads to a mean measurement uncertainty of 3.9\,ms$^{-1}$, including photon noise and wavelength calibration errors. The photon-noise uncertainty was estimated as described in Appendix~\ref{uncertainty}. External systematic errors of 2\,ms$^{-1}$ (spectrograph drift uncertainty) and of 4\,ms$^{-1}$ (guiding errors) were quadratically 
added to the mean measurement uncertainty (Boisse et al. 2010a). The mean error bar 
on each RV measurement is then about 6\,ms$^{-1}$. The RV 
measurements of HD\,109\,246 are listed in Table~\ref{table_rv}, available at the CDS. 
Table~\ref{table_rv} contains in its cols. 1-3, the time of the observation (barycentric
Julian date), the RV and its error, respectively.%the error bar on each RV measurement, respectively.

%__________________________________________________________________

\section{Stellar properties}
\label{stellar}
HD\,109\,246 (HIP\,61177) is a G0 dwarf with an apparent V-band magnitude of m$_{V}$\,=\,8.77 (SIMBAD database) and an astrometric parallax of $\pi$\,=\,15.14\,$\pm$\,0.68\,mas measured by Hipparcos (van Leeuwen 2007). Allende Prieto \& Lambert (1999) derived a stellar mass M$_{*}$\,=\,1.01\,$\pm$\,0.11\,M$_{\odot}$ and a radius R$_{*}$\,=\,1.02\,$\pm$\,0.07\,R$_{\odot}$. 

We determined the star's effective temperature, gravity and metallicity using the spectroscopic analysis of Santos et al. (2004). This analysis was performed on
the high-SNR average spectrum obtained by summing all the SOPHIE spectra. The 
spectroscopic analysis gave the following results: an effective temperature T$_{\rm eff}$\,=\,5844\,$\pm$\,21\,K, a surface gravity log\,$g$\,=\,4.46\,$\pm$\,0.19, a micro-turbulence velocity V$_{t}$\,=\,1.01\,$\pm$\,0.02\,\,km\,s$^{-1}$, and a stellar metallicity [Fe/H]\,=\,0.10\,$\pm$\,0.05 dex. When combined with isochrones (da Silva et al. 2006)\footnote{Web interface available on http://stev.oapd.inaf.it/cgi-bin/param.}, these parameters 
yield a stellar mass M$_{*}$\,=\,1.04\,$\pm$\,0.10\,M$_{\odot}$, in agreement with the Allende Prieto \& Lambert (1999) value.
We also computed the projected 
rotational velocity and the stellar metallicity from the SOPHIE CCF as described in Appendix~\ref{CCF}. 
This gave $v \sin i$\,=\,3\,$\pm$\,1\,km\,s$^{-1}$ and [Fe/H]\,=\,0.14\,$\pm$\,0.09. This alternative estimation of the stellar metallicity is 
 consistent with the accurate determination based on spectral analysis.

We estimated the stellar activity level from the emission in the core of the 
\ion{Ca}{II}~H\&K bands measured on each spectra of HD\,109\,246. The
calibration used to derive $\log$\,$R'_\mathrm{HK}$ activity indexes from SOPHIE
spectra is described in Appendix~\ref{activityIndex}. Applied to HD\,109\,246, this calibration 
yields $\log$\,$R'_\mathrm{HK}$\,=\,-5.05\,$\pm$\,0.1, which  
translates into an RV dispersion that would be %of about 
smaller than 1\,ms$^{-1}$ according to our experience on HARPS. 
%According to the calibrations by Noyes et al. (1984) and Mamajek \& Hillenbrand (2008), 
%the $\log$\,$R'_\mathrm{HK}$ value of HD\,109246 corresponds to a rotation period 
%P$_{rot}$\,$\approx$\,25 %($\pm$$\sim$3) 
%days. This estimation is somewhat inconsistent
%with the $v \sin i$ value, which implies P$_{rot}$ $\leqslant$ 17 %($\pm$$\sim$1) 
%days. 
%The discrepancy may be explained by Mamajek \& Hillenbrand (2008) remark that 
%stars with $\log$\,$R'_\mathrm{HK}$$\leqslant$ -5.0 appear to have a poor correlation 
%between $\log$\,$R'_\mathrm{HK}$ and the Rossby number, from which P$_{rot}$ is derived. 
The stellar parameters adopted for HD\,109\,246 are listed in Table~\ref{param_star}.

%---------------------------
\begin{table}
  \centering 
  \caption{Stellar parameters for HD\,109\,246. (1): Parameter derived from the SOPHIE CCF (see Appendices). (2): Parameter derived from the SOPHIE spectra (see Appendices).}
  \label{param_star}
\begin{tabular}{lc}
\hline
\hline
Parameters   & Values  \\
\hline
Sp. T.              &   G0V \\
m$_{V}$             &  8.77\\
B - V                &  0.64  \\
$\pi$     [mas]  &   15.24\,$\pm$\,0.68\\
T$_{eff}$    [K]  & 5844\,$\pm$\,21 $^{(2)}$   \\
log g    [cgs]  &    4.46\,$\pm$\,0.19 $^{(2)}$\\
Fe/H  [dex]  &   0.10\,$\pm$\,0.05 $^{(2)}$\\
v $\sin$i    [\,km\,s$^{-1}$]  & 3\,$\pm$\,1 $^{(1)}$  \\
%P$_{rot}$ & [d] & $\leqslant$ $\sim$18 \\
M$_{*}$    [M$_{\odot}$]  & 1.01\,$\pm$\,0.11   \\
R$_{*}$    [R$_{\odot}$]  & 1.02\,$\pm$\,0.07   \\
$\log$\,$R'_\mathrm{HK}$   [dex]  &  -5.05\,$\pm$\,0.1 $^{(2)}$\\
%Age  &  [Gyr] &  x\\
Distance  [pc] & 65.6 \\
\hline
\end{tabular}
\end{table}

%__________________________________________________________________

\section{HD\,109\,246b's orbital solution}
\label{orbite}

	\subsection{Evidence for a planetary companion}

As shown in Fig. 1, the RV data of HD\,109\,246 exhibit a peak-to-peak variation of $\sim$95\,m\,s$^{-1}$ and a dispersion of $\sigma_{RV}$\,=28\,m\,s$^{-1}$. A clear signal with a semi-amplitude larger than 30 ms$^{-1}$ is identified in the RV periodogram around 68 days. The dispersion of the bisector span values (BIS) is relatively low, with $\sigma_{BIS}$\,=9.9\,m\,s$^{-1}$ 
(Fig.~\ref{BIS}). No periodicity is detected in the bisector periodogram and no correlation is present in the bisector-RV plot (Fig.~\ref{BIS}). All these observations support a 
scenario where the RV variations of HD\,109\,246 are caused by the gravitational 
perturbation of an orbiting companion.

%----------------------------------------------------------- 
   \begin{figure}[h]
   \centering
  \includegraphics[width=9cm]{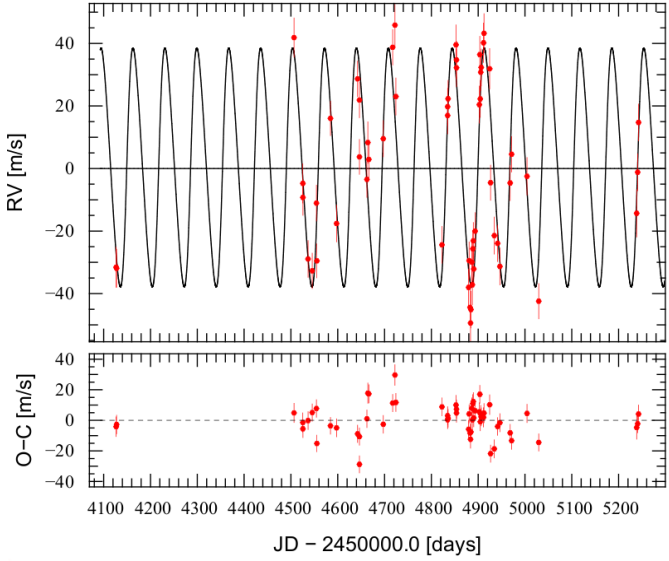}
      \caption{Radial velocities (\textit{top}) and residuals to the best-fit Keplerian model (\textit{bottom}) for HD\,109\,246 as a function of 
      barycentric Julian date. The best-fit Keplerian model is represented by the black curve. The residuals have a dispersion of 10.2\,m\,s$^{-1}$.
              }
         \label{bjdRV}
   \end{figure}
%--------------------------------------------------
%----------------------------------------------------------- 
   \begin{figure}[h]
   \centering
  \includegraphics[width=8.5cm]{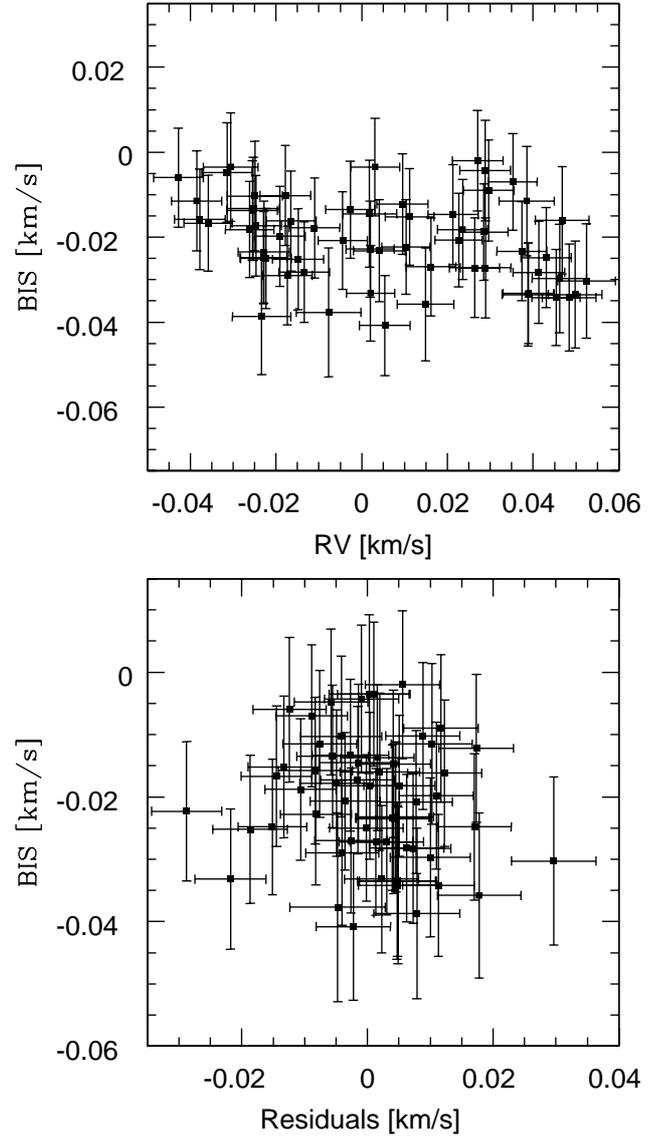}
      \caption{ Bisector span (BIS) as a function of RV 
      (\textit{top}), and as a function of the RV residuals 
      (\textit{bottom}). Uncertainties in the BIS are 
      assumed to be twice the RV ones. In both panels 
      the scale is the same in the x and y axes. No correlation is seen between BIS and RV residuals. The extra uncertainty seen in the RV data is unlikely to be caused by stellar activity.     
           }
         \label{BIS}
   \end{figure}
%-------------------------------------------------- 

%----------------------------------------------------------- 
   \begin{figure}
   \centering
  \includegraphics[width=8.5cm]{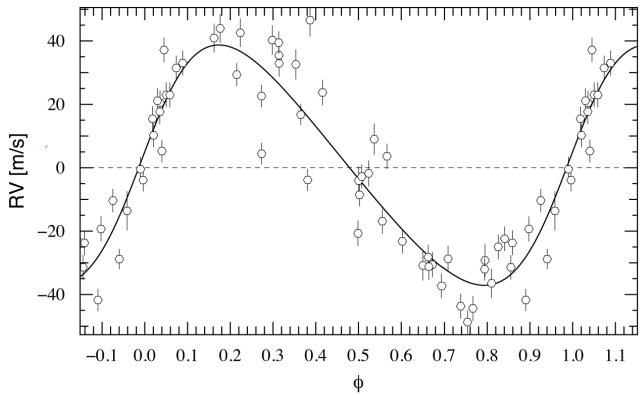}
      \caption{ Phase-folded RV curve for HD\,109\,246. The solid line
      represents the Keplerian fit with a reduced $\chi^{2}$ equal to 1.8. The period is 68.2\,d, the eccentricity is 0.17 and the semi-amplitude is 38.2\,m\,s$^{-1}$. Individual error bars are also plotted with a typical value of 6\,m\,s$^{-1}$. % and the colors indicate the measurement dates.
              }
         \label{phase}
   \end{figure}
%--------------------------------------------------
%--------------------------------------
\begin{table}[h]
  \centering 
  \caption{Keplerian solution and inferred planetary parameters. Results in the first column are based on the original RVs. Results in the second column are based on the corrected RVs (see text for details).}
  \label{param_p}
\begin{tabular}{lcc}
\hline
\hline
Parameters   & Original velocities & Corrected velocities \\
\hline
$VR$$_{mean}$  [\,km\,s$^{-1}$] &  -19.463 $\pm$ 0.004 & -19.464 $\pm$ 0.002 \\
$P$    [days]   &   68.20 $\pm$ 0.17 & 68.27 $\pm$ 0.13\\
$K$          [\,m\,s$^{-1}$]     &  38.2 $\pm$ 2.2 & 38.2 $\pm$ 1.6\\
$e$                &  0.17 $\pm$ 0.07  &  0.12 $\pm$ 0.04\\
$\omega$    [deg]  &   273 $\pm$ 25 & 235 $\pm$ 29 \\
$T$$_{0}$    [JD]  & 54831.7 $\pm$ 4.2   & 54824.6 $\pm$ 4.7 \\
$M$$_\mathrm{P}$ $\sin$ $i$ $^{1}$  [M$_{\rm Jup}$] &    0.76 $\pm$ 0.11 & 0.77 $\pm$ 0.09  \\
$a$$^{1}$   [AU]  &   0.33 $\pm$ 0.09  &  0.33 $\pm$ 0.08 \\
$\sigma_{(O-C)}$    [\,m\,s$^{-1}$]  &    10.2 & 7.7 \\
\hline
\end{tabular}
\begin{list}{}{}
\item[$^{1}$] assuming M$_{*}$\,=\,1.01\,$\pm$\,0.11\,M$_{\odot}$
\end{list}

\end{table}

%---------------------------------------------------

We fitted the RV data of HD\,109\,246 with a Keplerian model using a Levenberg-Marquardt algorithm, after selected starting values with a genetic algorithm. The best solution is a mildly 
eccentric orbit (e\,=\,0.17\,$\pm$\,0.07) with a period P\,=\,68.20\,$\pm$\,0.17 days 
and a semi-amplitude K\,=\,38.2\,$\pm$\,2.2\,ms$^{-1}$. This signal corresponds to a 
planet of minimum mass M$_{\rm p}$\,=\,0.76\,$\pm$\,0.11\,M$_{\rm Jup}$ orbiting 
HD\,109\,246 with a semi-major axis of 0.33$\pm$\,0.09 AU. The best-fit 
Keplerian model is plotted superimposed to the SOPHIE velocities in Fig.~\ref{bjdRV}. 
The corresponding phase-folded RV curve is shown in Fig.~\ref{phase} and 
the orbital elements of the planet are listed in 
Table ~\ref{param_p}. 
%Uncertainties in the orbital elements were determined from Monte Carlo experiments.
Uncertainties correspond to the 0.95 confidence interval after 5000 Monte-Carlo simulations. The tool used for the model fitting was successfully employed in exoplanetary surveys (e.g. Mayor et al. 2009; Bouchy et al. 2009b; H\'ebrard et al. 2010a).
The uncertainties in the minimum mass and in the semi-major axis take into account 
the uncertainties both in the stellar mass and in the signal semi-amplitude, period and 
eccentricity.

The residuals to the best-fit Keplerian model, $\sigma(O-C)$ = 10.2\,ms$^{-1}$, are 
a bit large compared to the mean error bar of 
$\approx$\,6\,ms$^{-1}$. The reduced $\chi^{2}$ is equal to 1.8, which suggests that an 
additional variation with a dispersion of $\sim$8\,ms$^{-1}$ is present 
in the data. If stellar jitter was the source of the excess variability (in spite of an estimated value lower than 1\,ms$^{-1}$, see Sect.~\ref{stellar}), we could expect a negative correlation between the bisector span and the RV residuals (e.g. Queloz et al. 2001; Boisse et al. 2010b).
%A CITER MON PAPIER SIMU IOTA SI ACCEPTER AVANT
 As shown in Fig.~\ref{BIS}, no such correlation is observed in the data. We note that the expected correlation would be very close to the detection limits but the level of activity needed to cause the RV variations is inconsistent with the $\log$\,$R'_\mathrm{HK}$ value.

%----------------------------------------------------------- 
   \begin{figure}
   \centering
  \includegraphics[width=8cm]{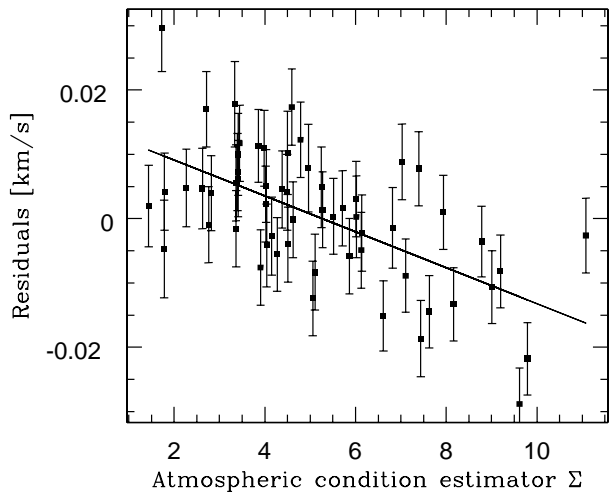}
      \caption{ 
      Residuals to the Keplerian fit as a function of the atmospheric condition 
estimator $\Sigma$, inversely proportional to seeing (see Sect.~\ref{spectrographIllumination} for the definition of $\Sigma$). The least-square fit represented by the solid line has a 
correlation coefficient of $-$0.6 and a Spearman coeficient of $-$0.56.
              }
         \label{res_seeing}
   \end{figure}
%--------------------------------------------------

About a quarter of gaseous giant planet are in a multiple planetary system. The excess variability may thus alternatively be due to the presence of an additional planet in the system. We did not find any indication for a second planet in the present data set. With a maximum semi-amplitude of 
14\,ms$^{-1}$, the RV residuals exclude the presence of an inner planet with $m_{p}$\,$\sin$\,i\,$\geqslant$\,0.28\,M$_{\rm Jup}$. A longer period planet could not induce a linear drift larger than 0.4\,ms$^{-1}$. Additional data will be needed to probe the presence of low-mass inner planets and to constrain the presence of possible outer companions.

The third and best possible explanation to the relatively large RV residuals is the 
underestimation of an instrumental effect. 
Described in the next Section, the so-called seeing effect seems to affect the observations of HD\,109\,246 (Fig.~\ref{res_seeing}).

%Of particular concern is the fact 
%that under good seeing conditions the input beam at the entrance of SOPHIE is 
%center-illuminated (input image smaller than the $3^{\prime\prime}$ sky 
%acceptance of the optical fiber). As discussed by Boisse et al. (2010), for 
%observations taken in high-resolution mode this result in a systematic decrease 
%in measured RV values (see Sect.~\ref{spectrographIllumination}). This so-called seeing effect has been observed on a sample of SOPHIE targets (Fig.~5 from Boisse et al. 2010a) and we show in Fig.~\ref{res_seeing} that it seems to affect the observations of HD\,109246 as well. Indeed, the residuals to the Keplerian
%fit are correlated with the seeing estimator $\Sigma$, defined as the relative flux 
%per unit of exposure time (up until recently the seeing was not monitored on the 1.93-m
%telescope). 

      \subsection{Spectrograph illumination limitation}
	\label{spectrographIllumination}
	
The large residuals of HD\,109\,246 RVs to the best-fit Keplerian model seem well explained by seeing effect due to bad scrambling of one multimode fiber (Fig.~\ref{res_seeing}). This effect was presented in details in Boisse et al. (2010a). We summarize here the main issue for the SOPHIE HR mode.

SOPHIE is a fiber-fed spectrograph. The stellar light collected by the telescope is 
led to the instrument through a standard step-index multi-mode cylindrical optical 
fiber. They have good azimuthal scrambling but the radial one is not perfect. 
 In order to improve the scrambling ability of the fiber and to stabilize the 
  illumination, optical double scrambler are usually coupled to the fibers (as it is done for the SOPHIE HR 
  mode) (Brown 1990; Hunter \& Ramsey 1992).
Non-uniform illumination of the slit or output fiber at the spectrograph entrance 
(variations in seeing, focus and image shape) decreases the radial-velocity precision, 
due to non-uniform illumination inside the pupil of the spectrograph that lead, with 
optical aberrations to variations in the centroids of the stellar lines on the focal 
plane.

Our experience on SOPHIE has led to identify remaining RV 
 limitations due to the incomplete fiber scrambling. %non-uniform illumination of the spectrograph. 
 We simulated the 
 optical path in the SOPHIE spectrograph and observed  that variations of the slit, pupil or optical fiber 
 illumination are directly translated on the spectrum.
 The displacement of the spectrum on the CCD detector is function of the wavelength (or the position on the detector) and is proportional to the width of the input image at the entrance of the fiber.
 Because the effect is not symmetric along the spectral orders and not monitored by the 
 calibration lamp, the final computed RV varies and its variation is proportional to the input image width.

 The value of the seeing is not monitored by SOPHIE up until recently. We estimated the atmospheric condition by calculating 
 the relative flux by unit of exposure time: 
\begin{equation}
\Sigma = \frac{SNR^{2}}{T_{exp} 10^{-M_{V}/2.5}}, \textrm{or for the same star  } \Sigma = \frac{SNR^{2}}{T_{exp}}
\label{s}
\end{equation} 
with SNR, the signal to noise ratio of the spectra, T$_{exp}$ the exposure time of the 
measurement, M$_{V}$ the visual magnitude of the target. This parameter allows a relative estimation of the seeing since we cannot evaluate the atmospheric absorption. The atmospheric condition estimator $\Sigma$ is inversely proportional to the seeing value.

%The seeing improves when the 
%seeing estimation $\Sigma$ increases.
   %When the seeing is good ($\leqslant$3", 
 %the fiber width), the input beam at the entrance of the spectrograph is 
 %center-illuminated (in HR mode). 
% When the value of the seeing is smaller than the fiber diameter ($\leqslant$3"), the input beam at the entrance of the spectrograph is center-illuminated (in HR mode) since the input image is smaller than the fiber width. 
 The seeing effect is expected to affect the measurements since the input beam at the entrance of SOPHIE is center-illuminated (input image smaller than the $3^{\prime\prime}$ sky acceptance of the optical fiber). 
 In this case, simulations predict a systematic decrease in measured RV values, which is observed in the obtained data, as shown in Fig.~\ref{res_seeing} of this paper and for a sample of SOPHIE targets in Fig.~5 from 
Boisse et al. (2010a).
 Moreover, when the input image is smaller than the diameter of the fiber, we are more 
 sensitive to effect of guiding and centering system that introduce RV variations. These
  RV variations are quite random and add some noise, 
whereas the seeing effect is directly related to the optical path in the spectrograph 
and may be quantified. We are modeling these variations with our data in order to remove 
this noise with a software tool even if this correction is only relative due to our inability to estimate the absorption parameter. The seeing effect may reach peak-to-peak RV variations of 20\,ms$^{-1}$ and it is the main current limitation of the SOPHIE accuracy.  
A new guiding camera is now in operation on SOPHIE and allows a better guiding and 
centering and 
also a monitoring of the exact seeing. It will then allow us to refine this method of correction.
 Tests on square and octagonal section fibers to optimize the fibers scrambling are now 
 under development and SOPHIE will be used as a bench test to validate these new feed optics. 
 Soon, a new double scrambler will be mounted on SOPHIE with the goal to reduce this effect by at least a factor 10.

	\subsection{Corrected velocities and improved orbital solution}

\begin{figure}
   \centering
  \includegraphics[width=9cm]{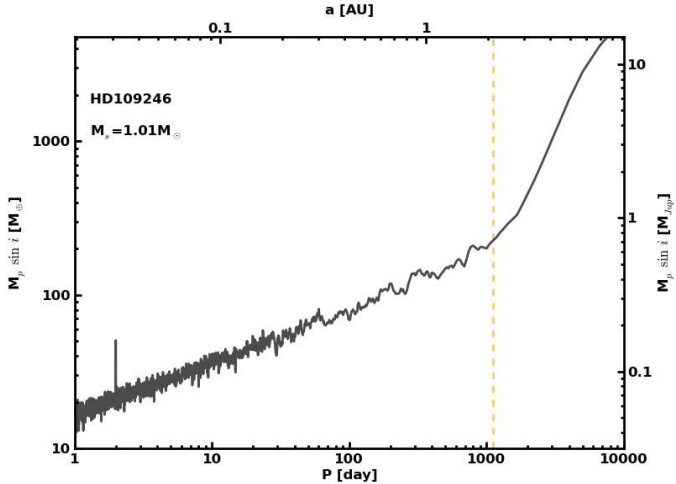}
      \caption{ Mass limit exclusion diagram with 99\% confidence level for an additional companion in the HD\,109\,246 system calculated from the residuals after the best-fit Keplerian model on the corrected velocities. The dashed line corresponds to the observation span. The two days peak expresses the lower efficiency at this period due to data sampling.
              }
         \label{exclu}
   \end{figure}

 In Fig.~\ref{res_seeing}, %an anti-correlation between the atmospheric condition estimator $\Sigma$ and the residuals around the Keplerian fit is detected. 
 the residuals to the Keplerian
fit are correlated with the atmospheric condition estimator $\Sigma$ (see Eq.~\ref{s}). %, defined as the relative flux 
%per unit of exposure time (up until recently the seeing was not monitored on the 1.93-m
%telescope). 
 We calculate the linear 
 correlation coefficient that is equal to -0.6, and the Spearman coefficient equal to 
 -0.56, that support the anti-correlation between the two parameters. The linear 
 function determined by a least-squares fit was subtracted from the radial-velocity 
 data measurements to correct for the seeing effect. 
%Following the method described in Appendix~D, we calibrated this effect and we corrected the RVs for it. 
Fitting a Keplerian model 
to the corrected RVs yields the orbital parameters reported in the right 
column of Table 3. The new planetary orbit is similar to the previous 
one; the major change introduced by the correction is to decrease 
the RV residuals to 7.7\,ms$^{-1}$. This new value does not change a lot our previous conclusions regarding the possible presence of an additional planet as illustrated in Fig.~\ref{exclu}. The remaining variability excess might be 
accounted for by an increase in guiding and centering errors when the input 
image is smaller than the fiber diameter (see Sect.~\ref{spectrographIllumination}) or by RV jitter due to stellar activity (see Sect.~\ref{stellar}). %We note that these new residuals do not change the limits for an inner or an outer companion.
%----------------------------------------------------------- 
%--------------------------------------------------

\section{Discussion and conclusion}

We report the detection of an extrasolar planet candidate around HD\,109\,246 
discovered with the SOPHIE spectrograph mounted on the 1.93-m telescope at 
the Haute-Provence Observatory. HD\,109\,246b is characterized by a
minimum mass of $0.77 \pm 0.09$ M$_{\rm Jup}$, an orbital period of 
$68.27 \pm 0.13$ days, and an eccentricity of $0.12 \pm 0.04$. The host star being slightly 
more metallic than the Sun, the discovery of HD\,109\,246b reinforces the correlation 
between giant planet occurrence and stellar metallicity (e.g., Santos et al. 2004; Santos et al. 2005; Fischer \& Valenti 2005). No photometric search for transits have been undertaken until now though the transit probability is slightly above 1\%.

The residuals to the single-planet Keplerian fit ($10.2$\,m\,s$^{-1}$) are larger than 
the mean error bar of $\approx$\,6\,m\,s$^{-1}$, which suggests the presence of an 
additional source of RV variability. We considered three possibilities: stellar 
jitter, the presence of an additional planet in the system, and an instrumental
effect related to seeing variations. We did not detect a clear signature of
stellar activity, nor that of an additional planet. On the other
hand, we identified a correlation between the residuals to the Keplerian fit and
a seeing indicator. Correcting the RVs of HD\,109\,246 for this systematic effect 
decreased the RV residuals to 7.7\,m\,s$^{-1}$ while preserving the original planetary orbit.
This result indicates that the seeing effect described in Boisse et al. (2010a) is 
at least partly responsible for the excess variability measured in the
velocities of 
HD\,109\,246. A new guiding camera is now in operation on the 1.93-m telescope at 
OHP. Besides providing better guiding and centering performances, this new camera 
allows for a direct monitoring of the seeing. Future observations shall thus 
allow the method described in this paper to correct SOPHIE data for 
the seeing effect to be refined. 
Moreover, a new double scrambler will be mounted on SOPHIE to improve the scrambling ability of the fibers and remove the seeing effect. 

\begin{figure}
   \centering
  \includegraphics[width=8cm]{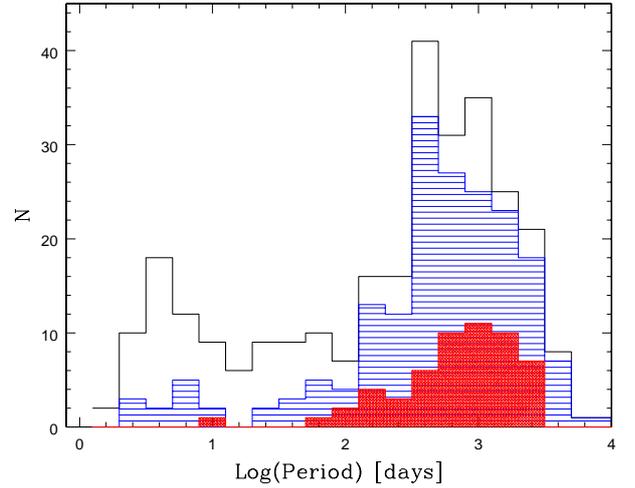}
      \caption{ Period distribution of the known extrasolar planet orbiting single dwarf stars discovered via RV surveys. With a period of 68 days, HD\,109\,246b is in the middle of the "period valley". Horizontal stripped illustrates the distribution for planets more massive than 0.8 M$_{Jup}$ and the shaded one for planets more massive than 4 M$_{Jup}$.
              }
         \label{histo}
   \end{figure}

The most remarkable property of HD\,109\,246b is that its orbital parameters 
place it within the so-called ``period valley'' in the period distribution 
of extrasolar planets (Fig.~\ref{histo}). The period valley extends from $\sim$10 to 
$\sim$100 days (Udry et al. 2003). It separates the pile-up of hot Jupiters 
from the population of giant planets with periods above $\sim$1 year.
HD\,109\,246b mass reinforces the fact that light planets ($m_{p}$\,$\sin$\,i\,$\leqslant$\,0.8\,M$_\mathrm{Jup}$) are preferentially found to orbit with short period ($P\,\leqslant\,100$d) (Udry \& Santos 2007). Ignoring multiple stars systems, RV surveys do not detect massive planets with $m_{p}$\,$\sin$\,i\,$\geqslant\,4$\,M$_\mathrm{Jup}$ with period less than 100 days. The two components left are HD\,162\,020b, a probable brown dwarf (Udry et al. 2002) and one of the two massive components orbiting HD\,168\,443 (Marcy et al. 2001; Udry et al. 2002). Including the transit surveys detections, planets more massive than 4\,M$_\mathrm{Jup}$ are not found in the ``period valley''. 

A comparison with the models of Mordasini et al. (2009) shows 
that HD\,109\,246b may belong to the population of ``main clump'' planets. %If so, it probably moved through type II migration from an inital formation site near 4--6 AU to its present location at $\sim$0.3 AU.
If so, it probably started to form near 4-6 AU and then moved inwards through type II migration. 
 As outlined by Mordasini
et al. (2009), the number of giant planets orbiting inside 1 AU 
is an important observational parameter for planet formation models. 
Additional detections of planets like HD\,109\,246b will thus help characterize the profile of the surface density of solids below 1 AU, and the efficiency with 
which migrating cores can accrete planetesimals.

%----------------------------------------------------------- 
%--------------------------------------------------

\begin{acknowledgements}
The authors thanks all the staff of Haute-Provence Observatory for their 
contribution to the success of the {\it SOPHIE} project and their support 
at the 1.93-m telescope. 
We wish to thank the ``Programme National de Plan\'etologie'' (PNP) of CNRS/INSU, the 
Swiss National Science Foundation, and the French National Research Agency (ANR-08-JCJC-0102-01 and 
ANR-NT05-4-44463) for their continuous support to our planet-search 
programs. AE is supported by a fellowship for advanced researchers from the Swiss
National Science Foundation (grant PA00P2\_126150/1). 
NCS would like to thank the support by the European Research Council/European Community under the FP7 through a Starting Grant, as well from Funda\c{c}\~ao para a Ci\^encia e a Tecnologia (FCT), Portugal, through a Ci\^encia\,2007 contract funded by FCT/MCTES (Portugal) and POPH/FSE (EC), and in the form of grants reference PTDC/CTE-AST/098528/2008 and PTDC/CTE-AST/098604/2008. DE is supported by CNES. This research has made use of the SIMBAD database and the VizieR catalogue access tool operated at CDS, France. The authors thanks the referee for his careful reading and judicious remarks.
  
\end{acknowledgements}

\appendix
%-------------------------
\section{Determination of the RV uncertainty from the 
{\it SOPHIE} Cross-Correlation Function}
\label{uncertainty}

The methodology to compute the photon-noise uncertainty of radial-velocity 
measurement was described by Bouchy et al. (2001). In this approach, 
the quality factor $Q$ represents the instrinsic RV-information content of a 
given spectrum (quality and richness of spectral lines). This $Q$ factor 
is independent of the flux and is computed on a reference spectrum considered as 
noise free. The RV uncertainty $\delta{V_{RMS}}$ is then computed using the 
following relation: 
\begin{equation}
\delta{V_{RMS}} = \frac{\mathrm{c}}{Q\,\sqrt{N_{e^-}}}\,,
\end{equation}
with c the speed of light and $N_{e^-}$ the total number of photoelectrons counted over the whole spectral 
range. In practice one does not always have a noise free reference spectra and one needs 
to estimate correctly the RV uncertainty on real observed spectra. 
On a noisy spectrum, the high frequency structures due to noise 
increase artificially the $Q$ factor and lead to an underestimated uncertainty 
for the RV. We estimate that the computation of the $Q$ factor is 
correct for spectra with S/N ratio per pixel higher than $\sim$200.
For lower SNR, $Q$ is overestimated and $\delta{V_{RMS}}$ is therefore underestimated. 
 
We adapted our methodology directly to the Cross-Correlation Function 
which corresponds to an average spectral line in velocity space at much higher S/N than individual 
lines and may be considered as noise free for the computation of the quality factor.
In that case $Q_{CCF}$ can be expressed by the relation:
\begin{equation}
Q_{CCF} = \frac{ \sqrt{\sum_{i}{ (\frac{\partial{CCF(i)}}{\partial{V(i)}})^2 / CCF^{2}_{noise}(i)}}}
{\sqrt{\sum_{i}{CCF(i)}}}.\sqrt{N_{scale}}\,,
\end{equation}
where $CCF(i)$ is the Cross-Correlation Function measured for the velocity $i$ and $CCF_{noise}(i)$ is the quadratic sum of the photon noise and the detector noise integrated 
inside the CCF mask holes for the velocity $i$. The correction factor $N_{scale}$ corresponds to 
the scale of the velocity step in detector pixel unit. The RV uncertainty $\delta{V_{RMS}}$ is then equal to:  
\begin{equation}
\delta{V_{RMS}} = \frac{1}{Q_{CCF}\,\sqrt{\sum{CCF(i)}}}\,.
\end{equation}

\begin{figure}
\centering
\includegraphics[width=8 cm]{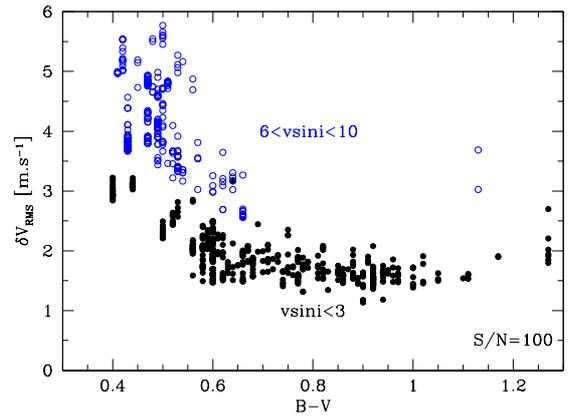}
   \caption{ RV uncertainty versus the index $B-V$ and for two classes of {$v$\,$\sin$\,$i$}, $\leqslant$3 {\,km\,s$^{-1}$} (plain dark circles) and in between 6 and 10 {\,km\,s$^{-1}$} (open blue circles), for SOPHIE observations in High Resolution mode with SNR of 100.  }
      \label{fignoise}
\end{figure}

This RV uncertainty, directly computed on the CCF, is robust and can be applied 
to any kind of spectra down to SNR of $\sim$20. Figure \ref{fignoise} shows the 
RV uncertainty scaled to an equivalent SNR of 100 and computed for a sub-sample 
of our planet search program with {\it SOPHIE} in HR mode. 
The RV uncertainty is plotted versus the index $B-V$ and for two classes of {$v$\,sin\,$i$} 
($\leqslant$3 {\,km\,s$^{-1}$} and in between 6 and 10 {\,km\,s$^{-1}$}). It illustrates the dependance with the 
effective temperature and the projected rotational velocity of the stars.

% AJOUTER EXEMPLE DE DVRMS a S/N=100 pour differents types spectraux. 

\section{Determination of {$v$\,sin\,$i$} and [Fe/H] from the 
{\it SOPHIE} Cross-Correlation Function}
\label{CCF}
We used the property that the CCF can be considered as an average line of the target spectrum. Line profiles are directly related to the stellar atmospheric parameters (e.g. abundances, thermal expansion), macro and micro-turbulence, rotation rate and instrumental profile. %The correlation mask include weak neutral lines. The profile of the CCF is fitted by a Gaussian whose parameters are FWHM, contrast and RV.
Then, stellar properties affect spectral lines included in the correlation mask and are reflected in the CCF. %sWe refer to Baranne et al. (1996) and Pepe et al. (2002) for a detailed description of the cross-correlation technique.\\
We can derive stellar properties in analyzing the FWHM and the contrast of the fitted Gaussian on the {\it SOPHIE} CCF. The $v \sin i$ of a star might be derived as described by Benz \& Mayor (1981) for the CORAVEL spectrograph. They refined their calibration afterward (Benz \& Mayor 1984). Queloz et al. (1998a) applied it to the ELODIE spectrograph. The stellar [Fe/H] calibration was first done by Mayor (1980) for the CORAVEL CCF, improved by Pont (1997). The both calibrations is done here for the SOPHIE spectrograph following the Santos et al. (2002) methodology %, refined by Santos et al. (2004),
 for the CORALIE spectrograph. The spectral line profile is dependent of the instrumental one, so these calibrations had to be derived for different instruments.

   \subsection{Calibration of the projected rotational velocity v sin i}
   \label{vsini}

We calibrated the 
relation between the {$v$\,sin\,$i$} and the width $\sigma$ of the CCF 
for both observing modes, High Efficiency (HE) and High Resolution (HR), 
of {\it SOPHIE} spectrograph using the cross-correlation masks G2 and K5 which have almost the same CCF width (negligible at the photon noise level). We note that the metallicity does not affect too much the width of the lines in comparison the the expected precision of our calibration.
For solar-type stars, these two variables can be related by:\\ 

\begin{equation}
\label{vsini_equa}
%K=214 \cdot \frac{m}{(m+M)^{2/3}} \cdot P^{-1/3}\,.
$v$\,\sin\,$i$=A \cdot \sqrt{\sigma^2-\sigma_0^2} 
\end{equation}

where $\sigma$  represents the measured Gaussian width of the CCF 
(FWHM/2$\sqrt{2\ln2}$), $\sigma_0$ the value of the expected $\sigma$ for 
a "non-rotator" of a given spectral type, and $A$ is 
a constant relating the ``excess'' width of lines to the actual 
projected rotational velocity $v$\,sin\,$i$. 

\begin{figure}
\centering
\includegraphics[width=8 cm]{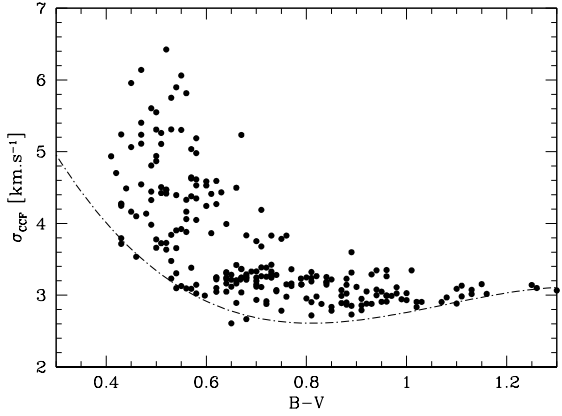}
   \caption{Gaussian width of the SOPHIE CCF versus $B-V$ index for 240 dwarf 
   stars observed in HR mode. The lower envelope determines the locus of "non-rotator" stars.}
      \label{figsigma}
\end{figure}

We first determine the quantity $\sigma_0$ by adjusting the lower envelope 
of the distribution of points in the diagram, shown in Figure \ref{figsigma} 
for HR mode, of $\sigma$ versus $B-V$ which is quite well related with 
stellar temperature for dwarfs with $B-V$ index in the range 0.4 - 1.3. 
We neglected here the effect of the metallicity. The adjusted lower envelopes 
for HR and HE mode are well described by:\\

\begin{equation}
\sigma_{0\_HR}=9.90-22.56(B-V)+22.37(B-V)^2-6.95(B-V)^3 
\end{equation}
\begin{equation}
\sigma_{0\_HE}=10.52-22.56(B-V)+22.37(B-V)^2-6.95(B-V)^3 
\end{equation}

\begin{figure}
\centering
\includegraphics[width=8 cm]{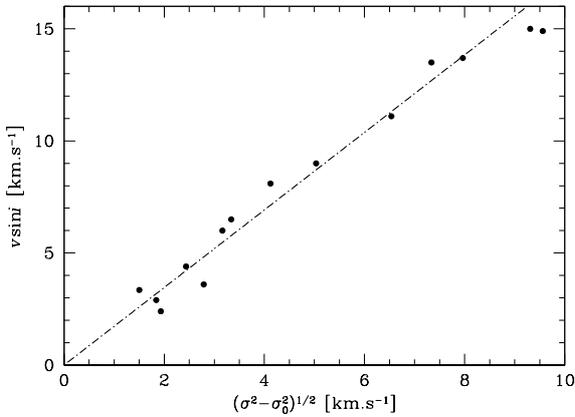}
   \caption{Known {$v$\,$\sin$\,$i$} versus measured $\sigma$ in the HR mode and the adjusted 
   slope.}
      \label{figvsini}
\end{figure}

The other variable in Eq.~\ref{vsini_equa} we need to determine is A, 
i.e., the constant relating the {$v$\,sin\,$i$} to the excess width of 
the CCF. For that we observed with {\it SOPHIE} 14 stars with known {$v$\,sin\,$i$} 
from the catalog of rotational velocities of F and G stars (Reiners \& 
Schmitt 2003) and we find A = 1.73 and A = 1.64 for respectively 
the HR and HE mode. Figure \ref{figvsini} shows the relation obtained 
between the known {$v$\,sin\,$i$} and the measured $\sigma$ for the HR mode. 
We note that this relation is valid up to {$v$\,sin\,$i$}=20\,km\,s$^{-1}$ (Bouvier, private communication). The estimated
uncertainty on {$v$\,sin\,$i$} is 1\,km\,s$^{-1}$.   

We find only three values of $v$\,sin\,$i$ from known planet hosting stars: HD189733 and HD80606 from Butler et al. (2006) and HD17156 from Fischer et al. (2007) with spectroscopic determination that may have its own uncertainty. Their estimations for these stars are about 1.5 kms$^{-1}$ above our estimate. On the other hand, our calibration find a value of 2.6 kms$^{-1}$ for the Sun $v$\,sin\,$i$. These values are satisfactory for the precision required.

The technique described above represents a simple 
way of obtaining an estimation of projected rotational velocities for 
dwarfs, simply as a by-product of the precise radial-velocity 
measurements. In particular, this calibration permits us to ob- 
tain quite easily values for the {$v$\,sin\,$i$} for all the stars in the 
{\it SOPHIE} exoplanet search program.

   \subsection{Calibration of the metallicity}
   \label{metallicity}
   
   The different SOPHIE masks are mainly built with neutral weak lines.
An important quantity of ferric ions increases the absorption coefficient for the lines of these elements. The lines then become deeper and the contrast of the CCF larger. But, the line contrast depends also on the spectral type, measured by the $B-V$ for a same luminosity class. The higher is the $B-V$, the lower is the effective temperature and the spectrum has more lines of metallic atoms. In addition, stellar rotation and Doppler shift, related to the relative speed of each atoms, modify the shape of the CCF widening and spreading the line profile but, it does not change the line surface. To take into account the stellar rotation and the macroturbulence at the stellar surface, we computed the total area $W$ of the Gaussian in spite of CCF contrast, $W$~=~$\sqrt{2\pi}($contrast$/100)\sigma$ %with $\sigma = FWHM/2\sqrt{2ln(2)} \approx FWHM/2.3548$ 
where $\sigma$ represents the measured Gaussian width of the CCF and the contrast is expressed in percentage, and studied its dependance as a function of the $B-V$ and metallicity. It is expected that the CCF surface grows with metallicity and with the $B-V$. 

A unique independent measurements of the metallicity has to be done for the calibration. We used a spectroscopic analysis based on a detailed study of each line (Santos et al. 2004). We determined accurately the [Fe/H] of 32 targets covering the $B-V$ and CCF surface domains. The mean uncertainty from the spectroscopic determination is about 0.1 dex. 

\begin{figure}[b]
   \centering
   \includegraphics[width=8.7cm]{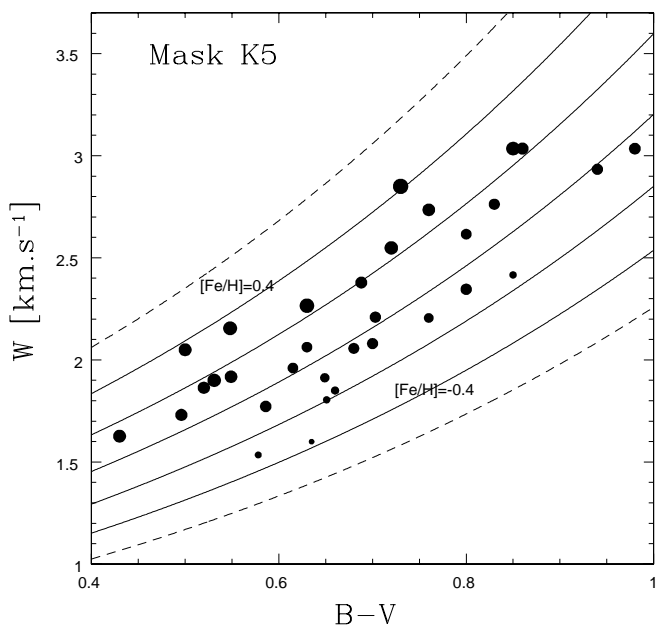}
      \caption{ Surface of the CCF, $W$, as a function of the $B-V$ for the K5 correlation mask. The dot size is proportional to the spectroscopic metallicity. The solid lines draw the relationship between the parameters for [Fe/H]=cst. The fit with an exponential is good in the domain of metallicity but it is obviously wrong for $B-V$$\geqslant$1.0.
                    }
         \label{calib}
   \end{figure}

%To do this calibration, we use SOPHIE data with CCF surface independent of SNR.  
{\it SOPHIE} has two modes of observations (Bouchy et al. 2009b), the High Resolution (HR) and the High Efficiency (HE) mode. Two fibers are used for each mode: one for the star and the other for the sky spectrum or for simultaneous calibration lamp exposure. The CCF surface is independent of the observation mode, provided that the SNR is sufficient (i.e. SNR $\geqslant$ 20 or SNR $\geqslant$ 100 for thorium simultaneous mode which produces an important background of scattered light all over the CCD). %, decreasing the accuracy on the CCF surface measurement). 
On the other hand, the CCF contrast, and so the CCF surface, depends on the calibration mask used. We studied  SOPHIE masks of spectral type G2 and K5 because the calibration is limited in $B-V$, excluding the coldest and the warmest stars. %A calibration is then done for each mask. 
For the K5 mask, the fit, shown in Fig.~\ref{calib}, gives:
\begin{equation}
[Fe/H] = 0.2615 + 3.9553 \log W - 2.2606(B-V)
\end{equation}
Samely, for the G2 mask, we obtained:
\begin{equation}
[Fe/H] = -0.9440 + 3.8807 \log W - 1.4992(B-V)
\end{equation}
These calibrations are valid for 0.43 $\leqslant$ ($B-V$) $\leqslant$ 0.98 and -0.47 $\leqslant$ [Fe/H] $\leqslant$ 0.44. % and xx $\leqslant$ xx $\leqslant$ xx.

\begin{figure}[h]
   \centering
   \includegraphics[width=8.7cm]{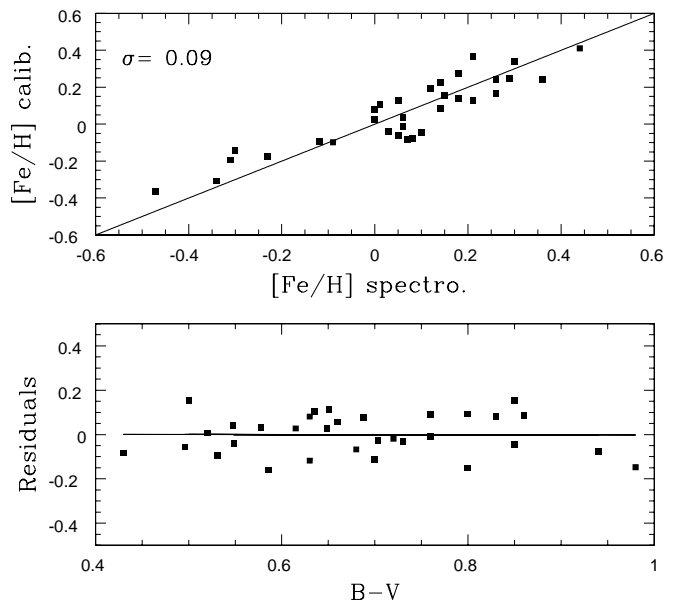}
      \caption{ \textit{Top}: [Fe/H] computed from the calibration as a function of the spectroscopic estimation. The solid line illustrates a 1:1 correlation. \textit{Bottom}: Residuals from the calibration as a function of the $B-V$.
                    }
         \label{resFEH}
   \end{figure}

\begin{figure}
   \centering
   \includegraphics[width=8cm]{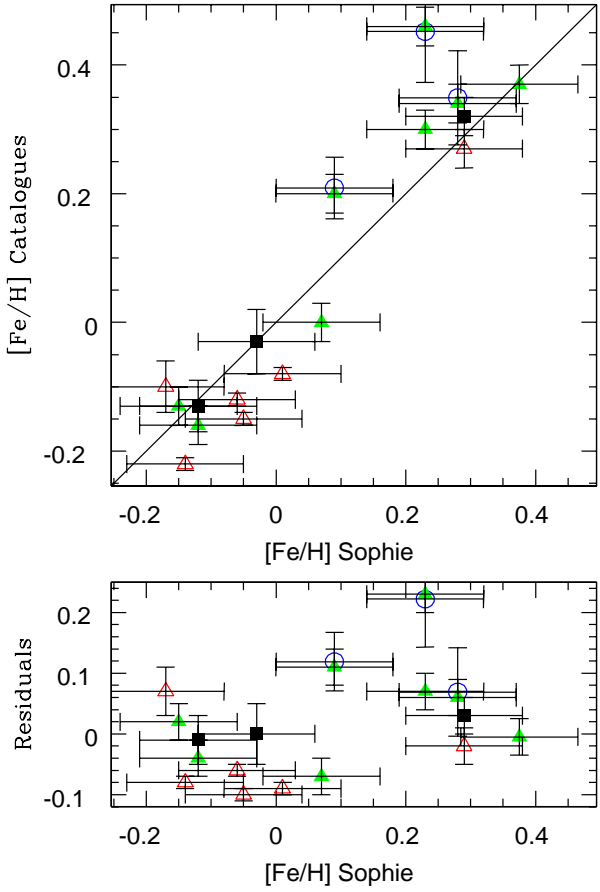}
      \caption{ \textit{Top}: [Fe/H] from others estimations from spectroscopic analyses: Sousa et al. (2006) (black squares), Sousa et al. (2008) (red open triangles), Gonzalez et al. (2007) (blue circle) and Fischer \& Valenti (2005) (filled green triangles) as function of those from SOPHIE calibration. The solid line illustrates a 1:1 correlation. \textit{Bottom}: Residuals from the top panel as a function of [Fe/H] from SOPHIE calibration. The rms of the residuals is 0.09 dex.
                    }
         \label{compFEH}
   \end{figure}

We compare the [Fe/H] values obtained from the CCF surface and from the spectroscopic analysis in Fig.~\ref{resFEH}. The dispersion of the residuals around the zero values is 0.09 dex which is satisfactory given the spectroscopic uncertainty. We do not observe any trend with the stellar temperature or $B-V$ as shown in Fig.~\ref{resFEH}. Our calibration is in agreement with others spectroscopic determination of the metallicity: Sousa et al. (2006), Sousa et al. (2008), Gonzalez et al. (2007) and Fischer \& Valenti (2005) as shown in Fig.~\ref{compFEH}.

%----------------------------------------------------------- 
%--------------------------------------------------
%----------------------------------------------------------- 
%--------------------------------------------------
	
%----------------------------------------------------------- 
%--------------------------------------------------

      %We collect the [Fe/H] estimation for the targets observed in the second {\it SOPHIE} subprogram planet-search sample (see Sect.~1). The histogram of the [Fe/H] computed for 848 stars observed is plotted on Fig.~\ref{histoFEH}. The sample has a total of $\sim$2000 targets, so the subsample of the observed star is not strictly without bias. As expected, the distribution is slightly higher (0.1 dex) compared to photometric [Fe/H] estimations for the solar neighborhood (Nordstr\"om et al. 2004; Holmberg et al. 2007).      
      
%----------------------------------------------------------- 
%   \begin{figure}
%   \centering
%   \includegraphics[width=8cm]{histo-feh-SP2-NewDRS_v1.eps}
%      \caption{ [Fe/H] distribution for 848 stars observed in the subprogram 2 of the SOPHIE planet-search sample. 
%                    }
%         \label{histoFEH}
%   \end{figure}
%--------------------------------------------------

The tool included in the {\it SOPHIE} data reduction software allows an immediate estimation of the [Fe/H] of the star with a good approximation ($\pm$ 0.09 dex). Our aim is to have a quick estimate of the target metallicity to focus our planet-search survey around over-metallic stars. One goal of the program is to find new hot-Jupiters known to be more prevalent around metal-rich stars (e.g., Santos et al. 2004; Santos et al. 2005).
 
%We compare our calibration with two others which samples have common stars with ours. %Some programs on SOPHIE measure targets already observed with ELODIE. A calibration based on the same analysis of the CCF were also done with ELODIE. For those stars observed by both instruments, the results of the calibrations according to each other is drown in Fig.xx. Both calibrations are consistent. There is a dispersion slightly larger for lower metallicities. This could be due to the fact that fewer stars with low metallicity are observed and, therefore, that the calibration has a greater uncertainty in this area. One can also compare our sample to those of the Nordstrom et al. (2004) calibration, revised in 2007 by Holmberg et al.. This calibration is derived from a photometric estimation. This method is known to have different biases.

%-------------------------

\section{Determination of $\log$\,$R'_\mathrm{HK}$ from the 
SOPHIE spectra}
\label{activityIndex}

\begin{figure}
   \centering
   \includegraphics[width=8cm]{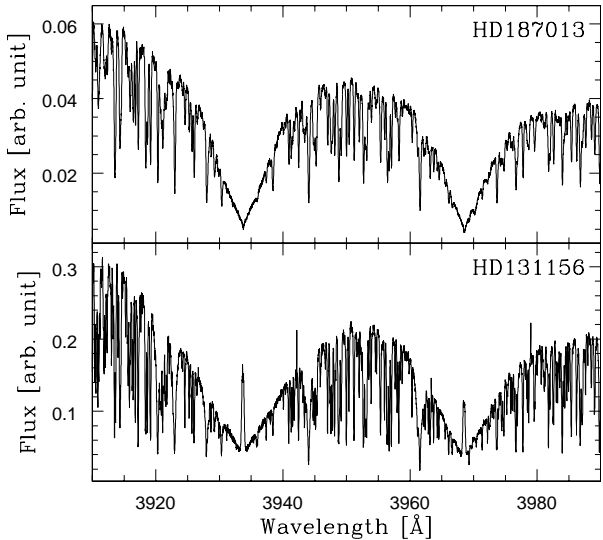}
      \caption{ Two SOPHIE spectra of the region of the CaII H and K lines between 3900 and 4000 \AA.  \textit{Bottom}: Active star. \textit{Top}:  Non active star.
                    }
         \label{spectre_Ca}
   \end{figure}

The SOPHIE spectra, with a bandpass from about 3900 \AA\ to 6800 \AA, include both Calcium II H and K resonant lines centered at 3968.49 \AA\ and 3933.68 \AA\ respectively. These lines are widely used indicators of stellar magnetic activity. In Fig.~\ref{spectre_Ca} are plotted two high-SNR SOPHIE spectra of the \ion{Ca}{II} lines central region for the chromospherically active star HD\,131156 and the inactive HD\,187013. An active star for which spots and plages are present on the photosphere is characterized by an emission in the center of the absorption line (e.g. Boisse et al. 2009).

%----------------------------------------------------------- 
%--------------------------------------------------

The stellar activity is parametrized by an index created by Wilson (1968) defined by the ratio between the emitted flux in the center of the lines and the continuum flux. 
The S-value is defined by the measure of the quotient of the flux in two triangular bandpasses centered on the H and K emission cores and two continuum regions on either side. We define our \ion {Ca} {II}~H\&K index following the Mt. Wilson S$_{\rm MW}$ index (Baliunas et al. 1995; Boisse et al. 2009):
\begin{equation}
\centering
Index=\frac{H+K}{B+V}
\end{equation}
 
 \noindent where H and K are the flux measured in the triangular 1.09 \AA\ FWHM window centered on each line of the \ion {Ca} {II} doublet, and B and V
 estimate the continuum on both sides with the flux measured in 20~\AA\ wide windows respectively centered on 3900~\AA\ and 4000~\AA. An automatic cut of the cosmic rays are done. Each line is located in two consecutive orders of the SOPHIE spectra. For the K line, we average the flux measured in two orders. As one of the H line is located on the edge of the an order where less flux is collected which introduce noise, only the flux measured in the H line of another order is kept. 

When studying the \ion {Ca} {II}~H$\&$K lines in SOPHIE spectra, one must account for the fact that they are located in a spectral region that suffers from contamination from background scattered light. %As the SNR is lower, the importance of the background subtraction increase. 
In order to use the signature of flux emitted in the \ion {Ca} {II} lines, we estimate a background light level and subtract it from the stellar spectra. For each mode, two fibers are used on SOPHIE, one of which is pointed on the target. The other fiber receives the sky spectrum or is fed with a Thorium-Argon lamp. The background light level of a stellar order is estimated from the same order of the other fiber by fitted a polynomial function on local minima. %Local minimum flux on pixel slices are fitted by a polynomial function to calculate the background level. 
 We put limits on the minimal SNR in the first order ($\lambda$$\sim$3955\AA) where both \ion {Ca} {II}~H$\&$K lines are found to have a good estimation of the activity index. %Boundary are fixed where no more dependency is observed between SNR and logR'$_{HK}$ values. 
 The boundaries are fixed where dependency between SNR and $\log$\,$R'_\mathrm{HK}$ is no longer observed. If the second aperture received the sky spectrum, the background light level is low and the minimal SNR is equal to 10. When the Thorium-Argon lamp feed the second fiber, the level of background light on the CCD is about $0.1\%$ of the stellar continuum and the minimal SNR to have a reliable value of $\log$\,$R'_\mathrm{HK}$ is fixed at 30. 
%These two SNR values is determined where there is no more dependency between SNR and logR'$_{HK}$ values. 
%The star spectrum is recorded in fiber A, while the ThAr lamp calibration spectrum is recorded simultaneously in fiber B. The calibration lamp produces a background of scattered
%light all over the CCD at the level of $0.1\%$ of the stellar continuum. 

\begin{figure}
   \centering
   \includegraphics[width=4.35cm]{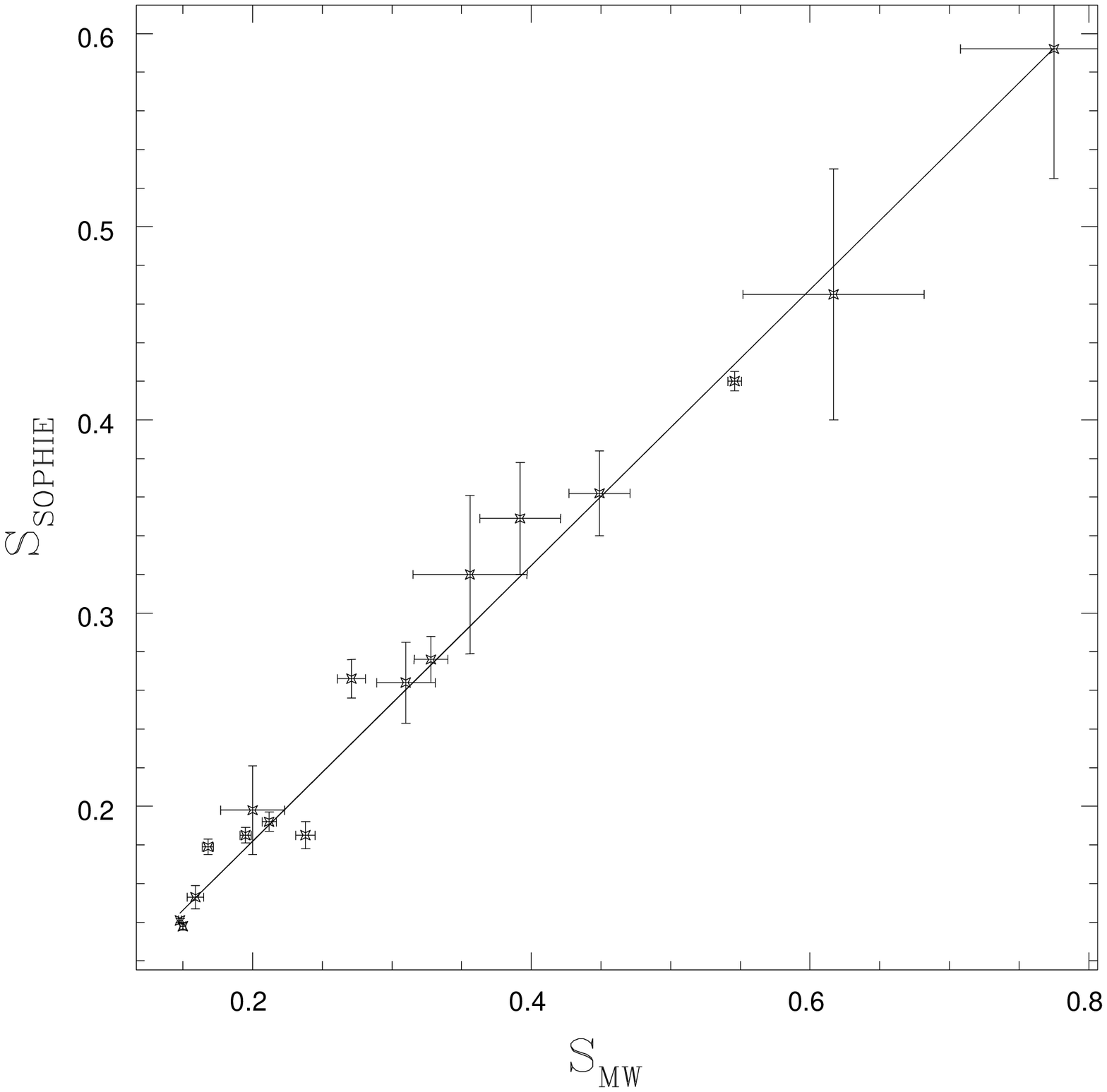}
    \includegraphics[width=4.35cm]{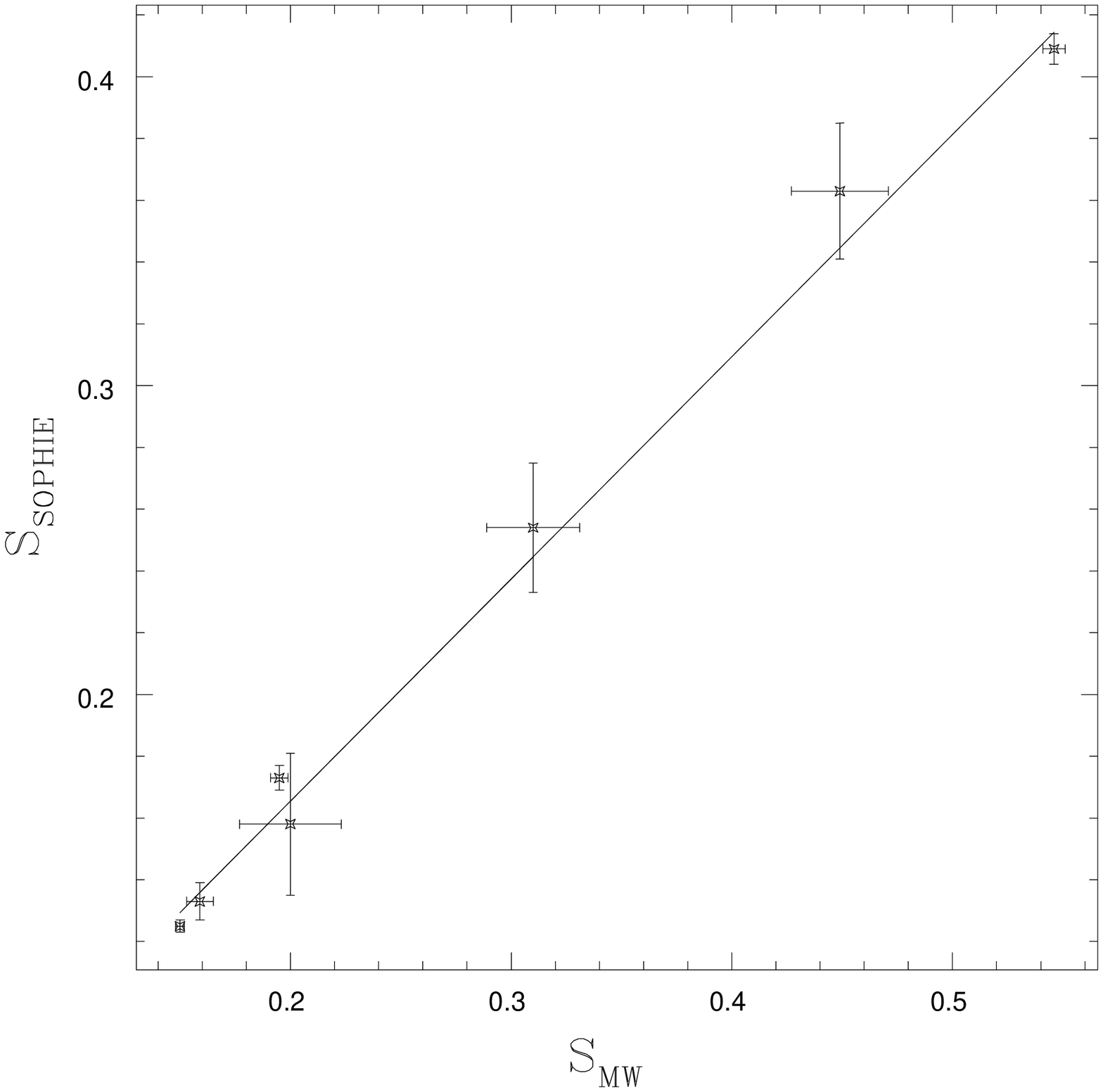}
      \caption{ Mean $S_{\rm SOPHIE}$ values as function of the Mount Wilson values from Duncan et al. (1991). The solid line is the best-linear fit to the data. \textit{Left:} High Resolution mode. (rms = 0.012) \textit{Right:} High Efficiency mode. (rms = 0.007)
                    }
         \label{calib_MW}
   \end{figure}

We measure a chromospheric index $S_{\rm SOPHIE}$ that is calibrated to the Mount Wilson index $S_{\rm MW}$ by observing calibration stars in HE and HR SOPHIE mode. The Mount Wilson values from Duncan et al. (1991) and our measurements $S_{\rm SOPHIE}$ are listed in Table~\ref{stars}, available in the electronic form of the paper. 
Table~\ref{stars} contains in its cols. 1-6, the name of the stars, the $S_{\rm MW}$ value and its error, the $S_{\rm SOPHIE}$ and its error and the SOPHIE observational mode, respectively.. The mean $S_{\rm SOPHIE}$ values compared with $S_{\rm MW}$ are shown in Fig.~\ref{calib_MW} for the set of calibrating stars for both mode of SOPHIE. In HR mode, the best linear fit to the data gives:
\begin{equation}
S_{\rm SOPHIE}\,=\,(0.71\,\pm\,0.03) S_{\rm MW}\,+\,(0.039\,\pm\,0.006) 
\end{equation}
In HE mode, the relation is:
\begin{equation}
S_{\rm SOPHIE}\,=\,(0.72\,\pm\,0.02) S_{\rm MW}\,+\,(0.021\,\pm\,0.006)
\end{equation}

With this index, we can calculate the litterature's index corrected from photospheric emission, $R'_\mathrm{HK}$ (Noyes et al. 1984). This measure is available only for 0.43 $\leqslant$ ($B-V$) $\leqslant$ 1.2.

%----------------------------------------------------------- 
%--------------------------------------------------

We estimate the uncertainty of our index ($\pm$0.1\,dex) with the dispersion on our values for a same star over one year. This dispersion may be due to intrinsic variations of the stellar activity. But, we consider that our main error comes from the subtraction of the background light.
%As the SNR is lower, the importance of the background subtraction increase. It is why we put limits on the minimal SNR to have a good estimation of the activity indicator. 

%----------------------------------------------------------- 
%   \begin{figure}
%   \centering
%   \includegraphics[width=8.3cm]{histo-Rhk-SP2.eps}
%      \caption{ $\log$\,$R'_\mathrm{HK}$ distribution for about 900 stars observed for the second subprogram of the SOPHIE planet-search sample.
%                    }
%         \label{histoRhk}
%   \end{figure}
%--------------------------------------------------

%We collect the $\log$\,$R'_\mathrm{HK}$ available for the targets observed in the SOPHIE SP2 planet-search sample (about 900 stars) and plot the histogram of the values on Fig.~\ref{histoRhk}. The distribution presents two peaks illustrating the Vaughan \& Preston gap (1980) with an inactive peak near $\log$\,$R'_\mathrm{HK}$=\,-5.0 and an active one at $\log$\,$R'_\mathrm{HK}$=\,-4.5. This histogram is in agreement with the Northern sample of the solar neighborhood observed at the Mount Wilson (Wright et al. 2004).

%Compared with an identical distribution of southern stars (Jenkins et al. 2008), the number of active stars in our sample is much more important.

 %The Keplerian fit of the corrected velocities are reported in Table~\ref{param_p}. 
% The residuals are reduced from 10.2 to 7.7\,m\,s$^{-1}$. The eccentricity slightly decrease but 
% become more significant. 

 %----------------------------------------------------------------
 % Table 1 available electronically only
\onltab{1}{
 \begin{table*}
  \caption{Radial velocities of HD\,109\,246\ measured with {\it SOPHIE}.}
  \label{table_rv}
\begin{tabular}{ccc}
\hline
\hline
BJD & RV & $\pm$$1\,\sigma$  \\
-2\,400\,000 & (km\,s$^{-1}$) & (km\,s$^{-1}$)   \\
\hline
54126.66704 & -19.4950 &  0.0064    \\
54127.61475 & -19.4953 &  0.0061    \\
54506.61615 & -19.4215 &  0.0063    \\
54525.49634 & -19.4682 &  0.0063    \\
54525.63339 & -19.4727 &  0.0057    \\
54536.58945 & -19.4923 &  0.0059    \\
54545.51957 & -19.4961 &  0.0057    \\
54554.47615 & -19.4745 &  0.0057    \\
54555.49364 & -19.4929 &  0.0054    \\
54584.45332 & -19.4474 &  0.0055    \\
54597.46134 & -19.4810 &  0.0059    \\
54642.41898 & -19.4347 &  0.0057    \\
54646.36523 & -19.4415 &  0.0057    \\
54646.37998 & -19.4597 &  0.0056    \\
54662.35377 & -19.4669 &  0.0058    \\
54664.35273 & -19.4551 &  0.0066    \\
54666.38121 & -19.4605 &  0.0059    \\
54697.32306 & -19.4539 &  0.0058    \\
54717.29649 & -19.4247 &  0.0057    \\
54722.28695 & -19.4175 &  0.0067    \\
54724.28200 &  -19.4404 &  0.0059    \\
54822.66825 & -19.4878 &  0.0059    \\
54834.69504 & -19.4464 &  0.0059    \\
54834.72295 & -19.4436 &  0.0059    \\
54835.70783 & -19.4411 &  0.0059    \\
54852.62557 & -19.4238 &  0.0064    \\
54853.69041 & -19.4287 &  0.0060    \\
54853.71034 & -19.4312 &  0.0061    \\
54879.60628 & -19.5014 &  0.0059    \\
54880.58603 & -19.4929 &  0.0060    \\
54882.61111 & -19.5078 &  0.0059    \\
54883.69930 & -19.5128 &  0.0058    \\
54884.55744 & -19.5085 &  0.0059    \\
54886.47681 & -19.4934 &  0.0068    \\
54887.52148 & -19.5006 &  0.0064    \\
54888.58770 & -19.4891 &  0.0059    \\
54889.61323 & -19.4865 &  0.0059    \\
54890.61440 & -19.4955 &  0.0058    \\
54893.48340 & -19.4834 &  0.0059    \\
54902.50858 & -19.4430 &  0.0059    \\
54903.54493 & -19.4270 &  0.0059    \\
54904.47549 & -19.4412 &  0.0059    \\
54905.48583 & -19.4326 &  0.0058    \\
54906.52068 & -19.4311 &  0.0059    \\
54911.52222 & -19.4232 &  0.0063    \\
54912.49518 & -19.4202 &  0.0063    \\
54924.57038 & -19.4315 &  0.0065    \\
54926.45707 & -19.4679 &  0.0056    \\
54934.48378 & -19.4848 &  0.0059    \\
54941.55035 & -19.4873 &  0.0058    \\
54946.35550 & -19.4947 &  0.0059    \\
54968.38892 & -19.4680 &  0.0057    \\
54971.38402 & -19.4588 &  0.0057    \\
55004.36914 & -19.4659 &  0.0060    \\
55029.35623 & -19.5058 &  0.0056    \\
55238.54345 & -19.4777 &  0.0076    \\
55240.69127 & -19.4646 &  0.0059    \\
55242.61572 & -19.4487 &  0.0059    \\
\hline
\end{tabular}
\end{table*}
}% end of onltab
%---------------------------------------------------------------------------------

%--------------------------------------------------
\onltab{2}{
\begin{table*}
  \centering 
  \caption{Stars used to calibrate the $S_{SOPHIE}$ in $S_{Mount\ Wilson}$.}
  \label{stars}
\begin{tabular}{cccccc}
\hline
\hline
Stars & $S_{MW}$ & $\sigma(S_{MW})$ & $S_{SOPHIE}$ & $\sigma(S_{SOPHIE})$ & Mode\\
\hline
HD\,97334  &   0.328   &  0.012    & 0.276 & 0.013 & HR\\
HD\,101501    &   0.310  &  0.021  &0.264 & 0.021 & HR\\
HD\,114378   &     0.238 & 0.007  & 0.185 & 0.007 & HR\\
HD\,114710   &  0.200   &  0.023  &0.198 &0.023  & HR\\
HD\,115404  &  0.546    &   0.005   &0.420 &0.005  & HR\\
HD\,126053 & 0.168    & 0.004  & 0.179 &0.004  & HR\\
HD\,131156 &  0.449   &   0.022   &0.362 &0.016  & HR\\
HD\,141004  &  0.159  &    0.006  & 0.153 &  0.003& HR\\
HD\,143761  &   0.148  & 0.001     &0.141 &0.027 & HR\\
HD\,149661    &      0.356 &   0.041 & 0.320 &0.004  & HR\\
HD\,152391   &     0.392        &     0.029 &0.349 &0.006  & HR\\
HD\,154417   &   0.271  &   0.010 & 0.266 &0.010  & HR\\
HD\,182101  &    0.212  &  0.005    & 0.192 &  0.001& HR\\
HD\,187013 &    0.150 & 0.002  & 0.138 & 0.002 & HR\\
HD\,190007 &   0.775  &    0.067  &0.592 &0.070  & HR\\
HD\,194012  & 0.195   &   0.004   & 0.185 &0.004 &  HR\\
HD\,201091  & 0.617   &  0.065    &0.465 &0.070 &  HR\\
\hline
HD\,101501   &    0.310     &0.021      & 0.254 &0.021  & HE\\
HD\,114710   &    0.200 & 0.023   &0.158 & 0.101 & HE\\
HD\,115404  &    0.546  &   0.005   &0.409 & 0.005 & HE\\
HD\,131156 &   0.449  &  0.022 &0.363 & 0.003 & HE\\
HD\,141004 & 0.159    & 0.006     &0.133 &0.002  & HE\\
HD\,187013  & 0.150   &  0.002    &0.125 &0.002 &  HE\\
HD\,194012  & 0.195   &  0.004    &0.173 &0.004 &  HE\\
\hline
\end{tabular}
\end{table*}
}
%---------------------------------------------------

\end{document}